\documentclass[superscriptaddress,english,aps,pra,twocolumn,showpacs,groupedaddress,floatfix]{revtex4}

\usepackage{epsfig}
\usepackage{graphicx} 
\usepackage{dcolumn}  
\usepackage{bm}       
\usepackage{color}
\usepackage{epsfig}
\usepackage{subfigure}
\usepackage{epstopdf}

\begin{document}
\title{Testable non-linearity through entanglement measurement}

\author{Wonmin Son}
\email{sonwm@physics.org}
\affiliation{Department of physics, Sogang University, Mapo-gu, Shinsu-dong, Seoul 121-742, Korea}

\begin{abstract}
A model of correlated particles described by a generalized probability theory is suggested whose dynamics is subject to a non-linear version of Schr\"odinger equation. Such equations arise in many different contexts, most notably in the proposals for the gravitationally induced collapse of wave function. Here, it is shown that the consequence of the connection demonstrates a possible deviation of the theory from the standard formulation of quantum mechanics in the probability prediction of experiments. The links are identified from the fact that the analytic solution of the equation is given by Dirichlet eigenvalues which can be expressed by generalized trigonometric function. Consequently, modified formulation of Born's rule is obtained by relating the event probability of the measuement to an arbitrary exponent of the modulus of the eigenvalue solution. Such system, which is subject to the non-linear dynamic equation, illustrates the violation of the Clauser-Hore-Shimony-Holt inequality proportional to the degree of the non-linearity as it can be tested by a real experiment. Depending upon the degree, it is found that the violation can go beyond Tsirelson bound $2\sqrt{2}$ and reaches to the value of nonlocal box. 
\end{abstract}
\maketitle

{\it Introduction -} Non-linear extensions of quantum dynamics have been suggested many times and it has been motivated from different perspectives \cite{Bassi13}. Ghirardi-Rimini-Weber (GRW) non-linearity \cite{Ghirardi86} was introduced to simulate an objective collapse of the wave function, although the precise mechanism of the collapse was still left as an open problem. The main motivation of the model is apparently evident such that we never observe macroscopic object (say of Planck's mass or larger) in the superposition of spatially distinct and separated states (say by a meter).

On the other hand, there are various gravitational proposals that lead to similar non-linearities as it is appeared in the GRW model \cite{Diosi87, Penrose96}. Some of these, however, are understood to be effective non-linearities such as the non-linearity in Gross-Pitaevski (GP) equation \cite{Gross61, Pitaevskii61} which merely arises due to the mean-field approximation applied to the linear many-body equation. In that case, the non-linearity is not fundamental but it can be taken as the derived quantity. Additionally, under the constraint of locality, a particular type of non-linearity, namely log non-linearity, had been suggested by Bialynicki-Birula-Mycielski (BBM) \cite{Birula76}. It is proved that the term still satisfies the condition of the separability between non-interacting sub-systems. With the log profiled non-linear interaction, it was proved that the existence of an isolated subsystem does not influence on the physical property of the other subsystems. An experimental test of the log non-linearity using the neutron interferometer showed that the effect can be considered significantly small \cite{Gahler81}.

In this paper, we show the different consequences of generalised theories proposing that the non-linearity in the Schr\"odinger equation allows different degrees of non-locality and the modification of the basic axiom in quantum theory. It is assumed that the property would not be an effective one as if it is induced by coupling to some external field as like the gravitational field. The mechanism of the non-linearity will not be immediate concern of us here, but the fact that the modified non-linear dynamic equation provides more general description of the probabilities than the original linear Schr\"odinger equation. The generality is identified through mapping between the non-linear equation and Bell violation under the generalized probability theory.

Additionally, it is shown that the consequence of the explored non-linearity is the modification of the Born rule in the standard quantum theory. The origin of this is, surprisingly, not dynamical in nature and, unlike the other proposals, is derived from the time-independent non-linear version of the Schr\"odinger equation. The result shows the clear connection between the different consequences of the non-linearity and provides us the instruction how to realise the non-local box in a real experimental setting. The parameter for the non-linearity also gives us the estimated precision in which the quantum mechanical prediction as a linear theory is considered to be accurate.

{\it Quantifying non-locality} - Before we state our main result, we give a brief introduction about Bell's inequalities for the local hidden variable model. Quantum correlations are the ``characteristic trait" of quantum state that, in fact, discriminate it from any classical theory \cite{Einstein35}. They also provide a resource for various forms of quantum information processing \cite{Nielsen00}. Specifically, quantum theory that the non-local character of a quantum system still complies with special relativity as per the quantum correlation predicted by Bell \cite{Bell} and Clauser-Horne-Shimony-Holt (CHSH) \cite{Clauser}. The Bell-CHSH  function is in the form of
\begin{equation}
\label{eq:Bell}
{\cal B}=E(\vec{a},\vec{b})+E(\vec{a},\vec{b}')+E(\vec{a}',\vec{b})-E(\vec{a}',\vec{b}')
\end{equation}
where $E(\vec{a},\vec{b})$ is a correlation function between two parties. In the local realistic (LR) model, the strict bound of the function is given as $|{\cal B}|\leq 2$ for the two outcome measurements. Statistically speaking, the model imposes a strong constraint on the joint probabilities given by the two classical dichotomic systems.

 Contrary to the local realistic prediction, the bound can be violated by a quantum mechanically correlated spin-$1/2$ systems. For a quantum system, $E_q(\vec{a},\vec{b})$ is given as the amount of correlations under the local measurements, parametrized by the three dimensional unit vectors  $\vec{a}$ and $\vec{b}$ at each site. Through an idealised quantification, the value becomes $E_q(\vec{a},\vec{b})=\vec{a}\cdot\vec{b}=\cos(\theta_{ab})$ for the maximally entangled state {\it e.g.} singlet state. Together with the provided quantum correlation, it can be shown that the maximal value of ${\cal B}$ goes up to the value $2\sqrt{2}$, called the Cirelson bound \cite{Cierlson}. Later, the physical origin of the maximal bound has been heavily discussed and become a source of intensive debate, see {\it e.g.}\cite{Epping13}. It is argued that there can be hypothetical theories that can achieve the highest value of Bell correlation, $4$, without contradicting causality. In our investigation, we have shown the possible links between the non-standard dynamical theory and the violation of Bell-CHSH inequality as the modification of fundamental axiom in quantum theory should be made.

We now present our main result about the Bell correlation in the generalized-probability framework and its relevant effect of non-linear Schr\"odinger equation on Born's rule. First of all, we start with non-linear equation where the single parameter potential of a system produces the function of particle density that gives rise to the non-linearity. After identifying the eigenstates of the differential equation, we show that the resulting probability does not coincide with what the standard Born's rule states on the derived quantity of the probabilities. Thus, the solution provides the generalization of Born rule to calculate Bell correlations in the non-linear regime and, consequently, show that any value of correlation between $2$ and $4$ can be attained depending on the degree of non-linearity that the dynamical system is subject to.

\begin{figure}[t]
\includegraphics[scale =0.42]{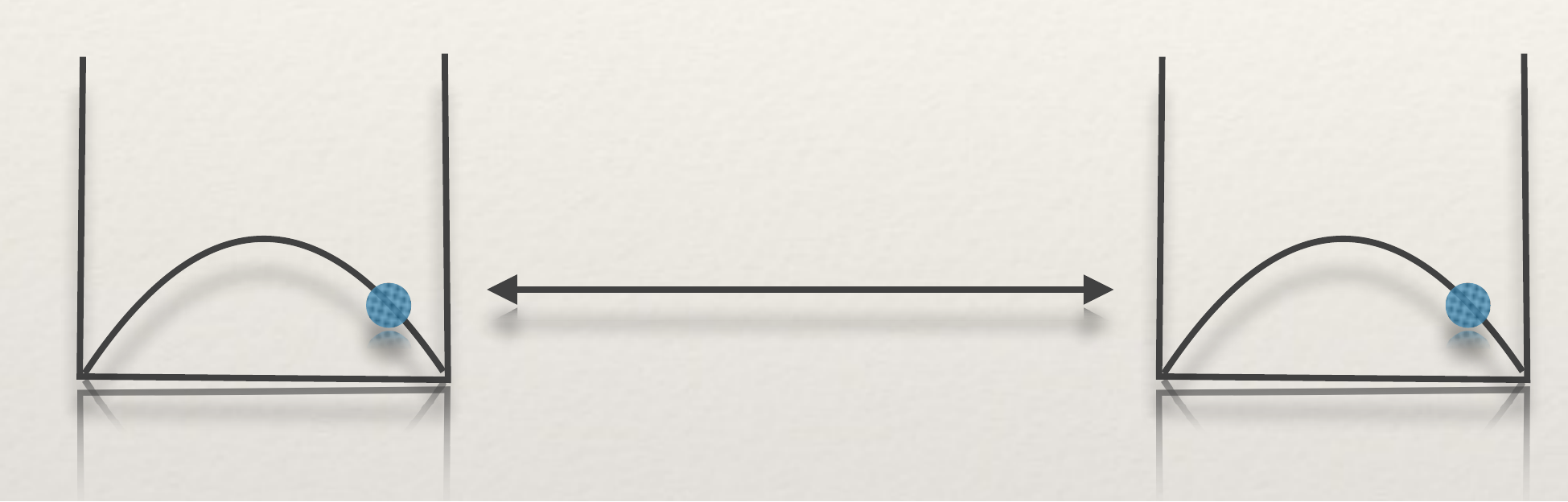}
\caption{Correlated particles which are confined in one dimensional box potentials}
\label{fig:Box}
\end{figure}

{\it One dimensional dynamic equation with non-linear interaction -}
Let us consider a dynamic equation of two particle systems that are confined in a non-linear wave potential well as it is sketched in FIG. \ref{fig:Box}. We assume that the dynamic equation in the coordinate of a relative motion is to be described by Schrodigner-like wave equation as 
\begin{equation}
i\hbar \frac{\partial}{\partial t} \psi(x,t)=\left(-\frac{\hbar^2}{2m}\frac{\partial^2}{\partial x^2}+g|\psi(x,t)|^{\epsilon}+V_{ext}\right)\psi(x,t)
\end{equation} 
when the parameter $x$ denote the relative position of the particles as $x=x_2-x_1$. Here, $\epsilon$ characterize the amount of non-linearity in the system. 

When $\epsilon=2$, the equation becomes a simplified version of non-linear Schr\"odinger equation which describes the density function of $N$ interacting Bosonic particles, called Gross-Pitowsky equation \cite{Gross61, Pitaevskii61}. The GP theory is a microscopic theory that describes the interacting non-uniform Bose gas at zero temperature {\it see e.g.}\cite{Leggett01}. It can also be related to non-equilibrium dynamics in cosmic structure and superfluid helium called Kibble-Zurek mechanism \cite{Kibble76, Zurek85}.  On the view of ordinary partial differential equation, it is just a second order differential equation (ODE) which has two parameters, spacial position $x$ and time $t$. 

In the other context of non-linear Schr\"odinger equation, the version of the equation also describes the situation when the particle is placed in a massive gravitational field. It is called Diosi-Penrose (DP) equation as it has been used to describe the mechanism for the wave function collapse model due to the gravitational field \cite{Diosi87, Penrose96}. In our current investigation, we address whether such a model can provide fundamentally different picture of nature then what can be derived from the standard quantum mechanics instead of exploring the actual physical consequences of the models by the non-linear effect.

As a matter of simplification, we consider a solution of the free evolution $\psi(x,t)=\phi(x) e^{-i\omega t}$ whose external potential matches to the resonant frequency as $V_{ext}=\hbar\omega$. In comparison to the original differential equation, the non-linear Schr\"odinger (NS) equation can take the form,
\begin{equation}
\label{eq:ns}
\frac{\hbar^2}{2m}\frac{\partial^2\phi(x)}{\partial x^2}+\bar{g}|\phi(x)|^{\epsilon}\phi(x)=0
\end{equation} 
where we have considered the case, $\bar{g}=-g$. With the parameters, the solution of the static NS equation is possible to be found analytically as 
\begin{equation}
\phi_{n,\alpha}(x)=\frac{\alpha T}{n \pi_{2,2+\epsilon}}\sin_{2,2+\epsilon}\left(n \pi_{2,2+\epsilon} x\right)
\end{equation}
when the positive constant of the coefficient takes the values $m\bar{g}/\hbar = (n \pi_{2,2+\epsilon})^4/|\alpha|^{2}$. Here, the special function $\sin_{2,2+\epsilon}(\cdot)$ denotes the generalized trigonometric function which is obtained in the work of Lundberg\cite{Lundberg79}  (see also\cite{Lindqvist95}), Levin \cite{Levin38} and \cite{Drabek99} as an analytical periodic function. The analytic form of the function is possible to be obtained as 
\begin{equation}
\sin_{2,2+\epsilon} (\theta) = F^{-1}_{2,2+\epsilon}(\theta)
\end{equation} 
where $F_{p,q}(\theta)=\int_1^{\theta} (1-t^q)^{-1/p} dt$ and $\pi_{p,q}:=2 F_{p,q}(1)$\footnote{$\pi_{p,q}=2\Gamma(1-1/p)\Gamma(1/q)/q\Gamma(1-1/p+1/q)$ with gamma function $\Gamma$, $\pi_p=\pi_{p,p}=2\pi/[p\sin(\pi/p)]$ and $\pi_2=\pi$}. The function $\sin_{p,q}(\theta)$ is a well-behaving periodic function whose periodicity is provided by newly defined irrational number $\pi_{p,q}$ as it becomes ordinary $\pi$ when $\pi_{2,2}=\pi$.

In the case of two particle system having additional degree of freedom, the spin-1/2 for example, the function $\phi(x)$ in Eq. (\ref{eq:ns}) can be interpreted as an object similar like wave function which gives the coincident probability of spin measurement for the correlated system. To be more specific, the statement can be represented as
\begin{equation}
p(a=b)=p(\uparrow\uparrow)+p(\downarrow\downarrow)=|\phi(x)|^2
\end{equation}
for the probability of the symmetric coincident events. Without detailed knowledge on the local wave function for the individual particles, the function $\phi(x)$ represents the probability density of the composite system that the particles register the coincident click. The only relevant external parameter for the function in this case will be the relative displacement that corresponds to the choices of measurement settings through the adjusted external parameters. 

The existence of the probability function also means that there is the {\it conjugate probability} and its corresponding wave function. The function can be obtained from the first derivative of the original wave function whose explicit form can be obtained analytically as $\cos_{2,2+\epsilon}(x)=\frac{\partial }{\partial x}\sin_{2,2+\epsilon}(x)=\bar{\phi}(x)$. Similarly to the ordinary trigonometric function, the conjugate probability of the non-linear Schr\"odinger equation can be obtained and it is given as
\begin{equation}
p(a\neq b)=p(\uparrow\downarrow)+p(\uparrow\downarrow)=|\bar{\phi}(x)|^{2+\epsilon}
\end{equation}
in order to satisfy the normalization condition. It is obtained from the constraint that the normalization condition of the probability is 
\begin{equation}
|\phi(x)|^2+|\bar{\phi}(x)|^{2+\epsilon}=1
\end{equation}
as it satisfies the condition of the generalized trigonometric function and $p(a=b)+p(a\neq b)=1$ from its definition. It implies that the solution of the wave equation is shifed from the original Born's rule as much as $\epsilon$ when one try to obtain the conjugate probability from the solution of the non-linear equation $\phi(x)$. In the next, together with the solution, we show that the deviation from Born's rule becomes a source of Bell violation beyond the Cirelson bound which is the maximum by the quantum state. 

\begin{figure}[t]
\includegraphics[scale =0.8]{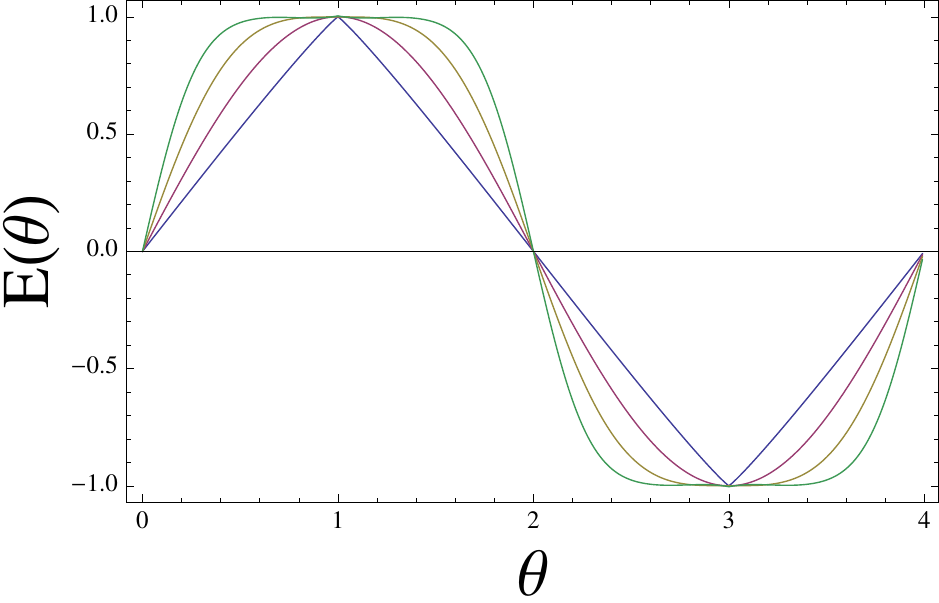}
\caption{Correlation function $E_{\epsilon}(\theta)$ as the nomalized parameter $2\theta/\pi$ varies. When $\epsilon=0$ the function becomes usual trigonometric function. As the value of $\epsilon$ changed, the convexsity of the function varies. When $\epsilon$ become large, the function approaches to linear function and when $\epsilon\rightarrow -1$, it becomes step function. }
\label{fig:correlation}
\end{figure}

In the setting for the Bell-CHSH inequality, the correlation functions with different measurements are required to be identified. They can be obtained from the solution of non-linear dynamic equation  as
\begin{eqnarray}
\label{eq:n-norm correlation}
E_{\epsilon}(\theta)=p(a=b)-p(a\neq b)
= |\phi(\theta)|^2 -|\bar{\phi}(\theta)|^{2+\epsilon}
\end{eqnarray}
where $\phi$ and $\bar{\phi}$ are the probability density specifying the measurement probabilities. As it is mentioned already, the second equation uses the normalization condition $|\phi(\theta)|^2 +|\bar{\phi}(\theta)|^{2+\epsilon}=1$. It also means that the correlation function is subject to a single value parametrization whose physical meaning is directly linked to the angle between the local measurement at the stations A and B. When the amount of nonlinearity $\epsilon$ is small, one can apply a possible Taylor expansion to the function as $|\bar{\phi}(\theta)|^{2+\epsilon}=|\bar{\phi}(\theta)|^{2}[1+\epsilon\log|\bar{\phi}(\theta)|+{\cal O}(\epsilon^2)]$ and then the correlation function becomes, 
\begin{equation}
\label{eq:nlcorr}
E_{\epsilon}(\theta)\approx (|\phi(\theta)|^2-|\bar{\phi}(\theta)|^2)-\epsilon |\bar{\phi}(\theta)|^{2}\log|\bar{\phi}(\theta)|
\end{equation}
where $\phi(\theta)=\sin_{2,2+\epsilon}(\theta)$ and $\bar{\phi}(\theta)=\cos_{2,2+\epsilon}(\theta)$.

\begin{figure}[t]
\includegraphics[scale =0.8]{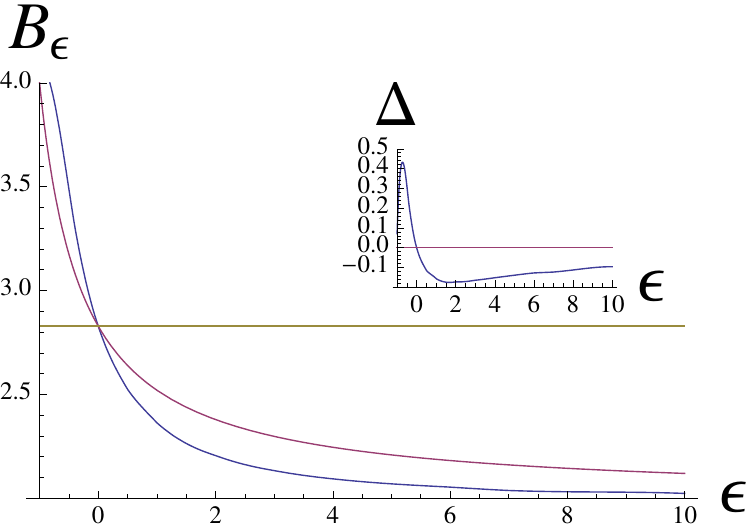}
\caption{Degree of Bell violation ${\cal B}_{\epsilon}$ for the case of different $\epsilon$. The numerical value of the Bell violation agree with the symmetric value of Bell correlation and it shows that there is continuous value $ 2<{\cal B}_{\epsilon}<4$ is obtained as the factor of modified Born's rule changed $-1<\epsilon<\infty$. The blue line is the value from the numerical simulation and the red line is plotted the function from our symetric conjecture.}
\label{fig:Violation}
\end{figure}

Interestingly, with the modified correlation function, one can show that Bell-CHSH inequality is violated beyond the Cirelson bound. The Bell function in (\ref{eq:Bell}) with the non-linear modification now becomes
\begin{equation}
{\cal B}_{\epsilon}=E_{\epsilon}(\theta_{ab})+E_{\epsilon}(\theta_{ab'})+E_{\epsilon}(\theta_{a'b})-E_{\epsilon}(\theta_{a'b'})
\end{equation}
where $E_{\epsilon}(\theta)$ is the same function appeared in (\ref{eq:nlcorr}). As it is discussed, when the Bell function becomes larger than 2, the system cannot be explained by local realistic model. It means that we need a theory outside realistic modelling as like the quantum theory in the regime. FIG. \ref{fig:Violation} shows the optimized value of ${\cal B}_{\epsilon}$ with respect to $\epsilon$. For the maximum value of Bell function, linear optimization over the parameters $(\theta_{ab},\theta_{ab'},\theta_{a'b},\theta_{a'b'})$ had been made and obtain the maximized value. The values with numerical optimization had been plotted in FIG. \ref{fig:Violation} represented by the blue curve. It is possible to find the value can be in the range of $2 < {\cal B}_{\epsilon} < 4$ as it is varied monotonically. It means that the Bell-CHSH correlation can take a value beyond the local realistic model as well as above the quantum theory depending upon the parameter $\epsilon$.

From the monotonic behaviour of the correlation, a conjecture is possible that the values of Bell function is varied according to the parameter $\epsilon$. As a rough estimation, the behaviour can be extrapolated from the optimised plot of Bell-CHSH function and a single parameterized ansatz of a simplified distribution can be given as 
\begin{equation}
\max [{\cal B}_{\epsilon}] =4/2^{(1+\epsilon)/(2+\epsilon)}.
\end{equation}
The behaviour has been drown as to fit that the value becomes 2 for $\epsilon\rightarrow\infty$, $2\sqrt{2}$ for $\epsilon=0$ and $4$ for $\epsilon\rightarrow -1$. The function can be taken as the extrapolated version of the optimised Bell function which resembles the behaviour of original function. The function of the Bell violation is plotted in Fig. \ref{fig:Violation} and it is fitted within marginal error that is shown by the inset of the figure for comparison. The functional behaviour follows the original one within the 10\% deviation from the actual optimized function.

In the figure, it is clearly shown that the Bell violation can be reached beyond the Cirelson bound when the non-linear parameter $\epsilon$ takes the negative value. That is the case when the non-linear term in the dynamical equation becomes inversely proportional to the wave function as the term asymptotically amplified at the limit of $\epsilon \rightarrow -1$. It is instructive that the value can swap all the region outside of local hidden variable model and reach to the maximum within the causality.

{\it Remarks} - It is shown that the model of correlated particles described by a generalized probability theory is suggested whose dynamics is subject to a non-linear version of Schr\"odinger equation. The consequence of the relation demonstrates a possible deviation from the quantum theory in the probability prediction of real experiments. The links are identified from the fact that the analytic solution of the equation is given by Dirichlet eigenvalue problem. Consequently, modified formulation of Born's rule is obtained by relating the event probability of measurement to an arbitrary exponent of the modulus of the eigenvalue solution. Such system, which is subject to the non-linear dynamic equation, illustrates the violation of the Clauser-Hore-Shimony-Holt inequality proportional to the degree of the non-linearity as it can be tested by a real experiment. Depending upon the degree, it is found that the violation can go beyond Tsirelson bound $2\sqrt{2}$ and reaches to the value of nonlocal box.

{\it Acknowledgement-} I would like to thank Prof. V. Vedral for his useful comments and suggestions. I also would like to acknowledge Oxford and KIAS for their hospitalities. 


\end{document}